\documentclass[12pt]{article}%
\usepackage{amsmath}
\usepackage{amsfonts}
\usepackage{amssymb}
\usepackage{graphicx}%
\begin{document}
%
\begin{titlepage}
\begin{flushright}
PUPT-2127\\
hep-th/0407209
\end{flushright}
\vspace{7 mm}
\begin{center}
\huge{Confinement and Liberation}
\end{center}
\vspace{10 mm}
\begin{center}
{\large A.M.~Polyakov\\
}
\vspace{3mm}
Joseph Henry Laboratories\\
Princeton University\\
Princeton, New Jersey 08544
\end{center}
\vspace{7mm}
\begin{center}
{\large Abstract}
\end{center}
\noindent
This is a review of topics which haunted me for the last 40 years,starting
with spontaneous symmetry breaking and ending with gauge/string/ space-time correspondence
While the first part of this article is mostly historical, the second contains some comments,
opinions and conjectures  which are new. This work is prepared for the volume
"Fifty Years of the Yang- Mills  Theories"
\vspace{7mm}
\begin{flushleft}
July 2004
\end{flushleft}
\end{titlepage}%

This article will discuss the subject which occupied most of my scientific
life - strong interaction of gauge fields. My first encounter with it happened
in 1964 when Sasha Migdal and myself (undergraduates at that time )
rediscovered the Higgs mechanism[ 1]. The idea of this work was given to us by
the remarkable condensed matter physicist , Anatoly Larkin. He said that in
superconductors there are no massless modes , presumably because of the
Coulomb interaction, and advised us to apply this to particle physics with
gauge fields. So we did $\ $and I still find some non-trivial elements in this
old paper. Experts in particle physics thought that our work was a complete
nonsense, but because of our age we were excused. However, it delayed the
publication of our paper for almost a year. Another year was taken by the
English translation of the JETP. As a result our work had no influence on
anyone except the authors.

I got a taste of field theory (so much despised by the particle theorists at
that time ) during this work and I liked it. I applied it to the theory of
critical phenomena, discovering operator product expansions [ 2] (here , as
turned out , I was behind Kadanoff and Wilson)and conformal symmetry of the
critical points [ 3](where I was the first) . Then I decided to study deep
inelastic scattering and $e^{+}e^{-}$ annihilation using the same methods ,
assuming that at short distances we have a conformal field theory [ 4]. I
found that the Bjorken scaling must be broken in a specific way - the moments
of the structure functions must scale according to the renormalization group .
More over, it was shown in these papers that the particles are produced in
jets by cascading process. They also contained what is called now "Altarelli
-Parisi equation" and "KNO\ scaling". I met David Gross in Kiev in the summer
of 1970 and discussed my formulae with him. He said that all they show is that
field theory has nothing to do with Nature since the Bjorken scaling is
clearly exact. In a few years he was plugging the asymptotically free
couplings in these formulae to display the (small) deviations from this scaling!

By 1972 it was clear that the renormalized coupling at small distances must be
either small or zero. Unfortunately I had a wrong Ward identity, showing that
it can't happen in the gauge theory and as an act of desperation was looking
at $\varphi^{4}$ theory with the wrong sign of coupling, stabilized by the
$\varphi^{6}$ term. The same idea was published later by Symanzcik, while I
never intended to publish, partly because of very negative reaction of my
colleagues ( who were right this time).

And then in the spring of 1973 Larkin brought the news from the US that
t'Hooft, Gross , Wilzcek and Politzer discovered a different sign of the one
loop beta function in gauge theories. After several days of checks I was
convinced that the new era begins. I was well equipped from my previous work
to proceed with the perturbative analyses of gauge theories but that was
already a vieux jeu for me. I wanted to explore the strong coupling region.

One thing which was on my mind for a long time were the classical field
configurations. In 1969 we discussed with Larkin whether the Abrikosov
vortices are normal particles represented as poles of some Green functions. We
didn't make much progress but the question bothered me since then. The second
stimulus I received from Faddeev's talk on the quantum sine -gordon theory. In
this case solitons were particles. It was unclear to me , however, to what
extent it is related to the integrability of the model. After a while I
realized that the relation is generic. This line of thought soon led to the
discovery of t'Hooft -Polyakov monopole.

The other line of thought came from the desire to understand the infrared
physics of the gauge fields. I received a strong stimulus from the work of
Vadim Berezinsky [ 5]. He explored two dimensional magnets and superfluidity
and had realized that the breakdown of the long ranged order in these systems
is caused by the condensation of vortices; analogously the 2d crystals are
melting through the condensation of dislocations. He developed a very complete
theory, starting from the lattice formulations of these problems. His results
were later rediscovered by Koesterlitz and Thouless.

I thought that the above picture of "dislocations" in space-time should solve
the infrared problem in gauge theory. I had in my disposal Berezinsky's
dissertation (to which I was a referee) - a complete and well written theory
of lattice systems in 2d (unfortunately it was never published and I have lost
my copy; Vadim died in 1980 at the age of 45). I started to generalize his
work to the gauge case and soon arrived at the non-abelian lattice gauge
theories. I was not in any hurry to publish it , thinking that no one could be
working in this direction and no one has a help from the Berezinsky
dissertation. This complacency was punished. Ken Wilson didn't need any help !
When I received his preprint I experienced a shock. His paper apart from the
things I already knew contained the criterion for confinement in the form of
the "area law". My consolation was that the idea of " dislocations" was still
not touched. And so I wrote a paper [ 6] which solved the problem of
confinement in the compact abelian theory. The basic ideas were as following.
Let us begin with the Maxwell action%
\begin{equation}
S=\frac{1}{4e^{2}}\int dxF_{\mu\nu}^{2}%
\end{equation}
with $F_{\mu\nu}=\partial_{\mu}$ $A_{\nu}-\partial_{\nu}A_{\nu}$ . Let us
assume that the vector potential is an angular variable. That means either
that the U(1) gauge group appeared as an unbroken part of a non-abelian group
or that in the lattice formulation the action is a periodic function of
$A_{\mu}.$In practical terms that means that the configuration space of $A-s$
must contain the fields of arbitrary number of magnetic monopoles. This is
analogous to the consideration of a quantum particle on a circle. It has the
same action as a particle on a line , but arbitrary windings must be included
into the functional integral. Another analogy is the real dislocations in a
crystal. In elasticity theory we can still use continuous displacement fields
but we have to add the singular fields coming from the dislocations and
reflecting the periodicity of the crystal.

The Wilson loop in this theory has the form%
\begin{equation}
W(C)=\int DAe^{-S(A)}\exp i%
{\displaystyle\oint\limits_{C}}
A_{\mu}dx_{\mu}=W_{0}(C)W_{M}(C)
\end{equation}
where the first factor comes from a simple Gaussian integral over $A$ while
the second represents the contribution of the monopoles. At this point it is
important distinguish the cases of three and four space-time dimensions. In
the former case the "dislocations" $($ or "instantons " as they were later
called by t'Hooft) are the point-like magnetic poles . Such an object (located
at the origin) is associated with the field strength%
\begin{equation}
F_{\mu\nu}\sim\frac{1}{e}\epsilon_{\mu\nu\lambda}\frac{x_{\lambda}}{x^{3}}%
\end{equation}
Correspondingly, its classical action is $\frac{const}{e^{2}}$ and the
contribution to the Wilson loop of the monopole located at $x$ is given by%
\begin{align}
W_{M}(x,C) &  \sim\exp(-\frac{const}{e^{2}})\exp i\eta(x,C)\\
\eta(x,C) &  =%
{\displaystyle\oint\limits_{C}}
A_{\mu}^{mon}(x-y)dy_{\mu}=\int_{S_{C}}\frac{(x-y)_{\mu}}{(x-y)^{3}}%
d^{2}\sigma_{\mu}(y)
\end{align}
It is clear that $\eta$ $(x,C)$ is a solid angle at which the contour is seen
from the monopole position and $S_{C}$ is an arbitrary surface bounded by the
contour $C$. The integral over the positions of the monopole will be dominated
by the configurations when the monopole is not too far from the loop and thus
we get the non- perturbative contribution to the Wilson loop of the size $R$
in the form $\exp(-\frac{const}{e^{2}})R^{2}$ . For large enough $R$ we must
sum over the plasma of randomly distributed monopoles. The monopoles form a
Coulomb plasma and as I knew since my '67 work on critical phenomena, it can
be reduced to the sine-gordon field theory. The sum over all monopoles can be
written as
\begin{equation}
W_{M}(C)\sim\int D\varphi\exp-e^{2}\int[\frac{1}{2}(\partial\varphi)^{2}%
+m^{2}(1-\cos(\varphi+\eta)]d^{3}x
\end{equation}
with $m^{2}\sim\exp-\frac{const}{e^{2}}.$ For large contours it is enough to
consider classical limit of the above theory, which is precisely the Debye
approximation. It is easy to see that the screening in this instanton plasma
gives confinement, Roughly speaking the one monopole contribution
exponentiates and generates the area law. More precisely one can easily solve
the classical equation coming from the action ( 6) since in the limit of large
flat contour the $\eta$ $(x,C)$ becomes a simple step function. It is
interesting that there exists a representation which combines the Gaussian and
the instanton parts together ( it was not described in [ 6] since I found it
later). Namely, it is easy to check that
\begin{align}
W(C) &  \sim\int DB_{\mu\nu}D\phi e^{-\Gamma}\\
\Gamma &  =\int d^{3}x[\frac{1}{4e^{2}}B_{\mu\nu}^{2}+i\phi\epsilon_{\mu
\nu\lambda}\partial_{\mu}B_{\nu\lambda}+\frac{m^{2}}{e^{2}}(1-\cos\phi
)]+i\int_{S_{C}}B_{\mu\nu}d^{2}\sigma_{\mu\nu}%
\end{align}
where we have introduced an independent antisymmetric field $B_{\mu\nu}.$ This
formula is very suggestive. The cosine term represents the monopoles. Without
the monopoles the field $\phi$ plays the role of the Lagrange multiplier and
the second term generates the Bianchi identity for the $B$ - field which
becomes an abelian field strength. The instantons here modify this Bianchy
identity (indeed, a single monopole gives a delta function in this identity)
and basically eliminate it in the infrared limit . A simple intuitive
explanation of the area law comes with the observation that if the Bianchi
constraint is dropped and $B$ becomes an independent field, the Gaussian
integral in ( 7) immediately gives the area law.

At the end of [6] I examined the four dimensional case. Strangely, most
readers missed this part , apparently thinking that the idea to go to four
dimensions never crossed my mind. It did.  In 4d the monopoles are particles,
localized in space but not in time. It was noticed in [ 6] that the only
non-perturbative effect (the instantons) comes from the monopole rings. In
terms of (7 ) , the Lagrange multiplier in 4d must be a vector. The classical
action of a ring of the length $L$ is proportional to $L,$ and thus the
instanton comes with the weight $\sim\exp(-\frac{const}{e^{2}}L).$The
contribution to the Wilson loop of the size $R$ comes from the monopole loops
with $L\sim R$ . For small charges the contribution is negligible. However,
the number of possible loops grows exponentially $\sim e^{constL}$ ( this is
the famous Peierls argument) and thus I predicted a phase transition to
confinement in the abelian 4d gauge theory. The analogue of ( 8) requires a
lattice regularization and can be written in the form
\begin{equation}
\Gamma=\int d^{4}x[\frac{1}{4e^{2}}B_{\mu\nu}^{2}+i\phi\wedge dB]+\frac{m^{2}%
}{e^{2}}\sum_{x,\mu}(1-\cos\phi_{x\mu})+\int_{S_{C}}B_{\mu\nu}d^{2}\sigma
_{\mu\nu}%
\end{equation}
Unlike the case of three dimensions, the $\phi$- field is massless at small
coupling and becomes massive after the phase transition to confinement. There
is a coy phrase in the paper "It is not clear whether this critical charge is
connected to the fine structure constant" . Alas, it is not.

A little later, t'Hooft and Mandelstam [7,8 ] published their views on the
abelian confinement in 4d. They started from the picture of "dual
superconductor" in which electric charges are dual to magnetic monopoles. If
the Higgs mechanism breaks conservation of electric charges, like it does in
superconductors, two magnetic charges will be confined by the Abrikosov vortex
connecting them. In the dual picture the Higgs field should describe magnetic
monopoles. At sufficiently large coupling they condense and as a result two
electric charges will be connected by an electric string.

This nice physical picture is completely equivalent to the one discussed above
( I don't remember if I understood it prior to reading [7,8 ] ; my diaries
don't contain it). The monopole loops condensation is precisely the Higgs
mechanism of t'Hooft and Mandelstam.

Next problem was to generalize it to the non-abelian case. Again the first
step is to find a classical solutions with finite action. This turned out to
be surprisingly easy [9 ]. Like in the case of the t'Hooft -Polyakov monopole,
the solution "solders" space- time and color space. That means that although
it breaks the Lorentz rotations $O(4)$ and color symmetry, it is still
invariant under a certain combination of both. It is hold together by its
non-trivial topology. Namely, consider a non-abelian gauge field $A_{\mu}(x)$
for which the field strength $F_{\mu\nu}$ goes to zero at infinity. This we
need to keep the action finite. The Euclidean space $R^{4}$ is bounded by a
sphere $S^{3},$and hence on this $S^{3}$ we must have asymptotically $A_{\mu
}=g^{-1}\partial_{\mu}g$ where $g$ belongs to the gauge group $G$ . Thus the
fields with the finite classical action are associated with the maps
$S^{3}\rightarrow G$ or the elements of the homotopy group $\pi_{3}(G)$=$Z.$
These integers $q$ are the values of the Chern classes and are expressed as
\begin{equation}
q=\frac{1}{16\pi^{2}}\int d^{4}xF\widetilde{F}%
\end{equation}
Finding of the solution is made easier by the self-duality equation which we
discovered in [ 9 ]. We "took a square root" of the Yang-Mills equations by
setting $F=\pm\widetilde{F}.$The Bianchi identity shows that any solution of
the self-duality is a solution of the Yang-Mills; self-duality turned out to
be quite interesting mathematically, leading to the new topological
invariants. More over, this solution is a true minimum of the Yang- Mills
action which can be written in the form
\begin{equation}
S\sim\int d^{4}x(F-\widetilde{F})^{2}+8\pi^{2}q\geq8\pi^{2}q
\end{equation}
The self-dual non-abelian instanton has many interesting properties. First of
all, it was obvious from the beginning that the classical solution in
imaginary time describes some kind of quantum mechanical tunneling.

Gribov ,t'Hooft, Callan, Dashen, Gross , Jackiw and Rebbi quickly made this
statement precise. Namely, take a gauge $A_{0}=0.$ Then the instanton solution
interpolates between various vacua of gauge theory in the following sense. In
the classical vacuum the field strength $F=0$ and $A_{n}(x)=g^{-1}\partial
_{n}$ $g.$The matrices $g(x)$ are separated into different classes defined by
the elements of $\pi_{3}(G).$Hence we have vacua labeled by the topological
charge $q$. The instanton solution has the property that $A_{n}(x,x^{0}%
=-\infty)=0$ and $A_{n}(x,x^{0}=+\infty)=g^{-1}\partial_{n}g(x),$ where $g(x)$
has topological charge $q=1.$Notice that precisely because of the fact that
$g(x)$ can not be continuously deformed to $I,$ the field of the instanton
must have non-zero field strength. More over, we should expect that the true
vacuum is a superposition of the above ones with the weight $e^{i\vartheta q}$
, where $\vartheta$ is a new physical parameter, analogous to the
quasimomentum in crystals. The same tunneling interpretation is applicable in
the case of the abelian instantons discussed above.

This was a nice interpretation, but the really stunning result came with the
work of t'Hooft [ 10 ]. He analyzed fermions in the field of instanton and
found that because of the zero modes, the instanton causes a dramatic symmetry
breaking. In the standard model this mechanism gives non-conservation of the
number of baryons! Instantons also solve the $U(1)$ problem of QCD , although
there are still some some puzzles in this case.

Finally, the presence of the $\vartheta$ angle introduces strong $CP$
violation in the theory since the topological charge is $CP$ odd. Why it is
not observed ? There are several possible explanations. My thoughts on the
subject is that there is a strong infrared screening of the $\vartheta$
-angle. I will have more to say on this subject below.

Let us return to the topic of the non-abelian confinement. Here the instantons
disappointed me. The problem is connected with the strong perturbative
fluctuations which potentially could obliterate the instanton. This is seen
from the one instanton contribution to the partition function of the $SU(N)$
gauge theory%
\begin{equation}
Z\sim\int d^{4}R\int\frac{d\rho}{\rho^{5}}(\mu\rho)^{\frac{11N}{3}}%
\end{equation}
In this formula $R$ is the position of the instanton, while $\rho$ its scale
(both are arbitrary parameters of the solution since the classical equations
are scale and translation invariant). The measure ( 12) has a very simple
meaning. The first two factors give a scale- invariant combination of $R$ and
$\rho$ while the last factor is related to the renormalized action on the
instanton, as can be seen from the relation%
\begin{equation}
\exp(-\frac{8\pi^{2}}{g^{2}(\rho)})=(\mu\rho)^{\frac{11N}{3}}%
\end{equation}
where $g(\rho)$ is the asymptotically free running coupling constant and the
expression in the exponential is the classical action of the instanton.

This semi-classical expression is valid if the exponential in (13 ) is small
or if $\mu\rho\ll1.$ Unfortunately the integral is dominated by the opposite
limit. So we have either to develop an approximate theory of the "instanton
liquid" , which was done by a number of people, or to hope ( as I initially
did) that some hidden symmetry protects the semiclassical approximation and
that the sum over instantons should generate confinement. This hope turned out
to be unrealistic in QCD, but in the case of gauge theories with
$\mathcal{N=}$ 2 supersymmetry it was justified by Seiberg and Witten twenty
years later. It is also interesting to notice that confinement in their model
(with the unbroken $U(1)$ gauge group) is precisely the one described above.

It was clear that we needed a more general approach in the non-abelian case.
Let us establish first some simple physical feature of confinement, Consider
the case of finite temperatures[11, 12 ] . As usual in this case we need to
integrate over the fields periodic in imaginary time with the period $\beta$
equal to the inverse temperature. However, if we try to fix the gauge
$A_{0}=0,$ we will have to use gauge transformations which are not periodic.
This is clear from the fact that the quantity $Tr$ $P\exp\int_{0}^{\beta}%
A_{0}(x,\tau)d\tau$ $\equiv Tr\Omega(x)$ is invariant under the legitimate
(periodic) gauge transformations and hence can't be eliminated. The partition
function can be written as%
\begin{align}
Z[\Omega]  &  =\int DA_{n}(x,\tau)\exp[-\int d^{3}xd\tau((\frac{\partial
A_{k}}{\partial\tau})^{2}+F_{kl}^{2})]\\
A_{k}(x,\beta)  &  =\Omega^{-1}A_{k}(x,0)\Omega(x)+\Omega^{-1}\partial
_{k}\Omega(x)
\end{align}
Here we managed to set $A_{0}=0.$ The price to pay is an extra (and important
) degree of freedom described by the time-independent matrix $\Omega.$ A
simple analyses shows that when we have a system of static charges located at
$x_{1}...x_{N}$ , their free energy is given by a correlation function%
\begin{equation}
\exp(-\beta F(x_{1}...x_{N})=\langle Tr\Omega(x_{1})...Tr\Omega(x_{N})\rangle
\end{equation}
where averages are taken with the measure defined by $Z[\Omega]$ and the
traces are taken in the fundamental representation of the gauge group. In the
confining phase the energy of a single quark should be infinite. That means
that $\langle Tr\Omega\rangle=0.$ As was pointed out in [ 11 ] the symmetry
which (if unbroken) ensures this condition is that of the \emph{center of the
gauge group. }Indeed, for the case of $SU(N)$ the measure $Z[\Omega]$ is
explicitly invariant under $\Omega\Rightarrow\exp(\frac{2\pi i}{N})\Omega$
(this symmetry reflects the fact that the gauge field itself is in the adjoint
representation and is insensitive to this transformation). At about the same
time t'Hooft also discussed the center of the group in a different approach,
based on the Kadanov-Ceva disorder variables. The heavy quarks represented by
the traces in (16 ) do change , however. More over, if the center symmetry is
unbroken and there is a mass gap, we have for quark and antiquark%
\begin{equation}
\langle Tr\Omega(x_{1})Tr\Omega^{\ast}(x_{2})\rangle\sim\exp(-M(\beta
)|x_{1}-x_{2}|
\end{equation}
which shows that the potential grows linearly with the distance. The center of
the group appeared because while the charges in the fundamental
representations are confined, the charges in the adjoint are not, being
screened by the gluons.

From this representation we can immediately conclude that at high temperature
(small $\beta$ ) the theory does not confine. The reason is that in this limit
there is not enough time to develop large $\Omega,$ and thus $Z[\Omega]$ will
be concentrated near $\Omega\approx I.$Actually, it is easy to show that if
$\Omega=I+\Phi,$ then $Z\approx\exp(-\frac{const}{\beta}\int d^{3}%
xTr(\nabla\Phi)^{2})$ . So, the center of the group symmetry is broken in this
limit but can be restored as we decrease the temperature. For example in the
abelian 3d model each instantonic monopole generates a vortex -like gauge
transformation $\Omega$ at $\tau=\infty.$ Random superposition of this gauges
restores the $U(1)$ symmetry and leads to confinement at zero temperatures.

Quark liberation can be understood in a very simple way by means of the
Peierls argument. Namely, while the energy of the string is proportional to
its length, the entropy of it also grows linearly (since the number of random
curves grows exponentially with their lengths). Thus at a certain temperature
the entropy takes over and infinitely long strings begin to dominate. That
means liberation.

So, the main prediction of [ 11, 12 ] was the existence of the quark - gluon
plasma after some temperature. It seems today that this phase is seen in the
experiments on the heavy ion collisions.

Around 1977 I started to feel that the semiclassical methods are insufficient
to solve the non-abelian confinement. A natural next step seemed to me the use
of loop variables and string theory. Indeed, the elementary excitations in the
confining vacuum are not point -like but string - like and the strings are
formed from the flux lines of color -electric fields. I decided to study the
equations in the loop space and to find a string theory which solves them [ 13 ].

Already in 1974, in his famous large $N$ paper, t'Hooft already tried to find
the string -gauge connections. His idea was that the lines of Feynman's
diagrams become dense in a certain sense and could be described as a 2d
surface. This is, however, very different from the picture of strings as flux
lines. Interestingly, even now people often don't distinguish between these
approaches. In fact, for the usual amplitudes Feynman's diagrams don't become
dense and the flux lines picture is an appropriate one. However there are
cases in which t'Hooft's mechanism is really working. This happens in the
$c\leq1$ matrix models in which the random surface is literally formed from
the dense lines of Feynman diagrams (as was shown by F. David and V. Kazakov
). Another case in which this mechanism may be at work are the matrix elements
with very large number of fields, like BMN\ operators. However in this case
some further clarifications are needed.

At the same time, one anticipation of the above paper holds in all cases - the
string interaction tends to zero as $N\rightarrow\infty.$ Therefore a great
simplification of the string picture is to be expected in this limit and
indeed occurs.

Another inspiring fact was the analogy with the 2d sigma models. In both cases
the theory is asymptotically free and develop a mass gap. This gap in the
sigma model is an analog of the non-zero string tension and thus confinement
in gauge theory. More over, the sigma models are completely integrable and
exactly solvable. That led me to the hope that there is something like
"integrability in the loop space" in gauge theories. To make it more concrete,
consider a field $\Psi(C)=P\exp%
{\displaystyle\oint\limits_{C}}
Adx$ , $W(C)=Tr\Psi(C).$ It is easy to check that the Yang-Mills equations are
equivalent to the following equations in the loop space%
\begin{equation}
\frac{\partial}{\partial x_{\mu}(s)}(\Psi^{-1}(C)\frac{\partial}{\partial
x_{\mu}(s)}\Psi(C))=0
\end{equation}
The "partials" here means the following (important) operation in the loop
space%
\begin{equation}
\frac{\delta}{\delta x_{\mu}(s_{1})}(A\frac{\delta}{\delta x_{\mu}%
(s)}B)=\delta(s-s_{1})\frac{\partial}{\partial x_{\mu}(s)}(A\frac{\partial
}{\partial x_{\mu}(s)}B)+...
\end{equation}
where the dots mean less singular terms. In terms of this operation the
classical Yang-Mills equation for the Wilson loop has the form
\begin{equation}
\frac{\partial^{2}}{\partial x_{\mu}^{2}(s)}W(C)=0
\end{equation}
These equations are classical. In quantum theory one expects contact terms on
the right hand side. These terms were a little later found by Makeenko and
Migdal and in the large $N$ limit they are remarkably simple. Classically, the
above equations are very similar to the ones of the non-linear sigma models
for the principal chiral field $\Psi(x)$ where $\Psi$ belongs to a Lie group.
In this case the equations are%
\begin{equation}
\frac{\partial}{\partial x_{\mu}}(\Psi^{-1}\frac{\partial}{\partial x_{\mu}%
}\Psi)=0
\end{equation}
These equations are known to be completely integrable by the Lax
representation. That led me to speculate that there should exist infinite
number of "loop currents" satisfying the equations
\begin{equation}
\frac{\partial}{\partial x_{\mu}(s)}J_{\mu}(s,C)=0
\end{equation}
as well as a Lax pair in the loop space. At present, 25 years later, elements
of complete integrability begin to appear, as we discuss below, although in a
somewhat different formulation.

It was clear that the loop space approach requires new string theory. We would
like to represent the Wilson loop as a sum over random surfaces $S_{C}$
bounded by the loop $C$%
\begin{equation}
W(C)=\sum_{S_{C}}e^{-F(S_{C})}%
\end{equation}
where $F$ is some unknown action. The natural choice of this action would be
the area of the surface (as was suggested by Nambu in the usual string
theory). Following Brink, di Vecchia, Howe and Deser and Zumino, it is
convenient to write it as a quadratic functional%
\begin{equation}
F=\int(\sqrt{g}g^{ab}\partial_{a}x\partial_{b}x+\mu\sqrt{g})+...d^{2}\xi
\end{equation}
where we do not write the fermionic terms (discovered by the above authors)
needed for superstrings. Here $g_{ab}$ should be treated as an independent
metric. Incidentally, the quadratic action was instrumental in solving the
Plateau problem by J. Douglas in the thirties. It is also crucial for quantization.

A surprise with this action is that in quantum theory it generates an extra
dimension. If we choose a conformal gauge $g_{ab}=e^{\varphi}\delta_{ab}$ the
"Liouville" field $\varphi$ drops from the first term of (24 ) making this
action Weyl invariant. However, after quantization it acquires a new life or,
which is the same, a non-trivial Lagrangian. It has the form%
\begin{equation}
L=\frac{26-D}{48\pi}((\partial\varphi)^{2}+\mu e^{\varphi})
\end{equation}
in a purely bosonic strings, while in the spinning string the critical number
26 is replaced by 10. This result implies that the natural habitat for the
random surface in D-dimensional $x$-space is D+1 dimensional ( $x,\varphi)$
space. The precise meaning of these words is that the wave functions of the
various string excitations depend on ( $x,\varphi).$ More over, the further
quantization of the $\varphi-$field leads to the conclusion that the metric in
this five dimensional (in case of QCD) space may be warped, having the form%
\begin{equation}
ds^{2}=d\varphi^{2}+a^{2}(\varphi)dx^{2}%
\end{equation}
The warp factor $a^{2}(\varphi)$ is determined from the condition of the
overall Weyl invariance of the theory. It can be interpreted as a running
string tension. I was helped here by the following analogy with the 2d systems
with the $SU(N)$ symmetries. Namely, the analogue of the spectrum of string
tensions of gauge theory is simply the mass spectrum of a 2d system (this is
obvious on the lattice in the strong coupling expansion). It is well known
that in the integrable 2d systems the typical mass spectrum is $m_{n}\sim
\sin(\frac{\pi n}{N})$ which becomes continuous as $N\rightarrow\infty.$ So, I
was not too shocked to conjecture the continuous spectrum of the string tensions.

This is not all, however. The equation (23 ) means that we are trying to
identify the wave functional of a string (the r.h.s.) with the Wilson loop of
gauge theory. In general the wave functional depends on the contour $C$ ,
parametrized by $x_{\mu}$ $=x_{\mu}(s).$This functional is invariant under
reparametrizations $s\Rightarrow\alpha(s)$ , provided that $\frac{d\alpha}%
{ds}>0.$But the Wilson loop, being defined by a contour integral, has larger
symmetry. It is insensitive to the change of sign of $\frac{d\alpha}{ds}$ or
to the backtracking of the contour (zigzag symmetry). The string theory,
therefore, must be such as to accommodate this property. This condition can be
formulated as follows. 

In string theory it is more convenient to discuss not the wave functionals but
the open string amplitudes, given by the expectation values of the vertex
operators defined at the boundary of the world disk. In the standard string
theory there is an infinite number of such vertex operators, corresponding to
the infinite number of the open string states. For example, the operator
$V(p)=\int ds\sqrt{h(s)}e^{ipx(s)}$ describes a tachyon, while $V_{\mu
}(p)=\int ds\frac{dx_{\mu}(s)}{ds}e^{ipx(s)}$ corresponds to a massless vector
state of the open string. Here $h(s)$ is the metric on the boundary of the
world disk.

It is important to notice that all vertex operators except the massless ones
depend explicitly on $h(s).$This dependence violates the zigzag symmetry
(roughly speaking, backtracking changes the length of the loop). Hence we must
be looking for a peculiar string theory in which there are infinite number of
closed string states and only finite number (corresponding to the massless
modes) of the open string states.

The key idea for solving this problem is based on warping [ 14]. Suppose that
the contour $C$ is placed at some position in the $\varphi$ space,
$\varphi=\varphi_{\ast}.$Then the masses of the open and closed string
excitations are related by%
\begin{equation}
M_{open}^{2}\sim a^{2}(\varphi_{\ast})M_{closed}^{2}%
\end{equation}
indicating a simple blue shift effect. If we place the contour at $\infty$ in
the $\varphi-$ space, where $a^{2}(\varphi_{\ast})=\infty,$ all massive open
string states disappear, while the massless remain. They are the "edge"
states of string which are dual to the states of the field theory. There is
also another, less convenient placement for the boundary ( T-dual to the
described above) but we will not use it here.

When reported at "Strings'97" these ideas were met with scepticism ("you keep
feeding us with beautiful mirages"-a reaction of one outstanding physicist).
That changed with the work of J. Maldacena who noticed that in the $N=4$
Yang-Mills theory, which is known to be conformally invariant and which was
already compared with supergravity by Ig. Klebanov, the isometries of the
metric ( 26) require it to represent AdS space of constant negative curvature
, that is fix $a^{2}(\varphi)\sim e^{\alpha\varphi}.$This example provided us
with an excellent theoretical laboratory.

The easiest case of this AdS/CFT correspondence is the limit of small
curvatures of the AdS space. It corresponds to the large Yang-Mills coupling
(which is our free parameter since the beta function is identically zero). In
this limit , instead of solving the sigma model directly, one can use the
method of effective action in the target space. Namely, it has long been known
in string theory, that the low-energy interactions can be obtained from the
supergravity action (of which we write only the relevant bosonic part)%
\begin{equation}
S=-\int d^{10}x\sqrt{G}e^{\Phi}(R+(\nabla\Phi)^{2}-|dB|^{2})-\sqrt{G}%
\sum|F_{p}|^{2})
\end{equation}
where $\Phi$ is a dilaton , $B_{\mu\nu}$ is an antisymmetric tensor, and
$F_{p}$ are various RR field strengths. It is almost obvious that there exists
a classical solution, representing $AdS_{5}\times S_{5}$, with constant
dilaton and zero $B-$field. Indeed, if we take the $F_{5}$ form, which is
self-dual, to be the volume form on the above 5d spaces, the last term in (
28) acts as a cosmological term, with the negative cosmological constant for
the first factor in the above product and the same but positive constant for
the second factor ( $S_{5}).$To extract the Yang-Mills correlation function we
have to follow the procedure discovered in [ 15,16]. It consists of several
simple steps. First, let us write the classical solution in the Poincare form%
\begin{equation}
ds^{2}=\sqrt{\lambda}(\frac{dx^{2}+dy^{2}}{y^{2}})+...
\end{equation}
where the dots represent the $S_{5}$ of the metric which is not important at
the moment (it represents extra scalar fields of the $N=4$ gauge theory), and
$\lambda$ , which determines the curvature of $AdS$ is related to the t' Hooft
coupling : $\lambda=g_{YM}^{2}N_{c}.$Each string excitation corresponds to a
certain operator in the gauge theory. Suppose that we look at a certain string
field $\phi_{n}(x,y)$ with the mass $M_{n}$ (at small curvature classification
of states in our string theory is of course the same as in the flat space). It
satisfies the wave equation%
\begin{equation}
(-\nabla^{2}+M_{n}^{2})\phi_{n}(x,y)=0
\end{equation}
A general solution of this equation has the following asymptotic behavior at
infinity ( $y\rightarrow0$ ), $\phi_{n}(x,y)\rightarrow y^{\Delta_{-}}%
\varphi_{n}(x),$where $\Delta_{n\pm}=2\pm\sqrt{4+M_{n}^{2}\lambda}.$The field
$\varphi_{n}(x)$ is conjugate to a certain operator $TrO$ $_{n}$ of gauge
theory (formed out of field strengths and covariant derivatives) in the
following sense. The generation function for these operators turns out to be
equal to the classical action as a functional of $\varphi_{n}(x)$%
\begin{equation}
\langle\exp N_{c}\sum\int dx\varphi_{n}(x)TrO_{n}\rangle_{YM}=\exp N_{c}%
^{2}S_{cl}[\varphi_{n}(x)]
\end{equation}
where we explicitly added the number of colors $N_{c}.$To calculate $S_{cl}$
we have to perturb the action (28 ) by the corresponding field $(.$e.g. for
the massless modes ,just to vary the fields already present in the action). If
we are interested in the two point functions on the gauge theory side, the
linear perturbation will suffice, otherwise non-linear terms will be needed.
Since the dimension of $\phi$ is zero, we conclude that the dimension of
$\varphi$ must be $\Delta_{n-}$ and hence the dimension of $Tr$ $O_{n}$ is
$\Delta_{n+}$. This can be trusted if $\lambda\gg1$ (the small curvature limit).

What these formulae tell us is that you have to solve non-linear classical
equations (the Einstein equations for the massless modes and string equations
for the massive ) and then extract the information about highly quantum regime
of the gauge theory. In some limited sense it looks as a realization of the
Einstein dream - to replace quantum theory by non-linear classical equation.

Another feature of this formula is that in a certain sense the theory of
gravity in D dimension is encoded by the Yang-Mills theory in D-1 dimension
"located" at infinity. It smells as a "holographic principle" proposed by
t'Hooft. On the other hand t'Hooft's argument was that when you put too much
energy into the system, black holes will be formed and their entropy is
proportional to the area and not the volume. I don't see any direct relation
of this argument to the above considerations. After all, the fact that we
describe gravity by the boundary fields $\varphi$ just amounts to the solution
of the Dirichlet problem and is rather prosaic. What is non-trivial, is that
the classical action ( or more generally the wave functional of the universe,
which in the semi-classical limit is given by the RHS of ( 31)) are related to
the Yang - Mills theory.

There has been a tremendous progress in associating various supergravity
solutions and their deformations with the gauge theories. However it seems
that perhaps it is time to leave supergravity alone. It is already abundantly
clear that it works. The real challenge is to go to the cases of large
curvatures. Here the results are more modest , but as I shall argue, quite promising.

There are several reasons to pursue these investigation. First, the problem of
quark confinement and of "QCD\ string" lie clearly beyond the supergravity
approximation, since the gauge coupling constant is running and becoming small
in the UV region. That means that the curvature of the 5d space varies and
becoming large at infinity. As a first step one can still be looking at the
conformal models, but in the cases when the curvature is large. The formalism
of effective action becomes quite useless in this case, and one has to attack
the sigma model directly.

One approach [17 ] is to consider the operators with the large quantum numbers
and to treat the sigma model semiclassically. The idea of tis approach is as
following. The lagrangian of the sigma model has the form%
\begin{equation}
S\sim\sqrt{\lambda}\int d^{2}\xi(\partial N)^{2}+...
\end{equation}
where $N$ is a hyperbolic unit vector (in the 6d Minkowski space), $N^{2}%
=-1.$The dots stand for the fermionic terms which neutralize the beta function
and make the model conformal on the world sheet. Various operators, formed of
the $N$ field and its derivatives acquire anomalous dimensions on the world
sheet $\delta=\delta(\lambda,\Delta,J,...),$where $\Delta$ is an anomalous
dimension in space time (which is one of the projections of the angular
momentum in the above 6d space), $J$ is the spin of the operator and the dots
stand for the other possible quantum numbers. The necessary condition for a
sigma model operator to describe a physical state in string theory is
$\delta=1.$ This relation determines the spectrum of the space time anomalous
dimensions $\Delta.$The above effective action calculation is equivalent to
using the one loop expression for $\delta$ , which is inadequate in general.

To get more interesting information, let us notice that the world sheet
dimensions can be viewed as eigenvalues of the sigma model hamiltonian,
provided that we put the theory on a cylinder. This hamiltonian has
eigenstates corresponding to small oscillation of the field $N.$ This is what
we have described above. But there are also a completely different states
corresponding to solitons. These solitons correspond to various classical
motions of the string. A good example is provided by the rotation of the
folded string in the AdS space. A simple classical computation gives the space
time anomalous dimension for the gauge theory operators with high spin $J$.
The result is $\Delta(J)-J=c(\lambda)\log J.$ Since such operators define deep
inelastic scattering, we come to a fascinating conclusion that this process in
a certain region can be viewed as exciting rotation of the folded string in
the warped 5d space!

Careful investigation of the various solitons led to the conclusion that the
spectrum of $\Delta$ coincides with the spectrum of an integrable
ferromagnetic chains[ 18 ]. The same conclusion follows from the study of
perturbation theory on the gauge side. In fact, the basic statements in the
case of weak gauge coupling has been known from the works by Lipatov [ 19] and
Faddeev and Korchemsky [ 20]. By now we see that the integrability of the
spectrum persists in the strong coupling limit. In general, it should reflect
complete integrability of the underlying sigma model. If we drop the fermionic
terms , as in ( 32), such integrability is well known on the classical level
[21] ; it was also checked that at the quantum level various anomalies don't
destroy it [ 22]. But what happens as we add the RR background to the
Lagrangian ? Several years ago I concluded that in the RNS formalism the RR
fluxes don't destroy integrability. The argument was as following. Let us add
the fluxes perturbatively. In the n-th order they result only in the change of
the boundary conditions for fermions, which must change sign when going around
n selected points. Now imagine that we are deriving conserved currents by
changing variables in the path integral.Since all currents contain even number
of fermions, the above change of the boundary condition should not interfere
with these changes of variables and thus the currents continue to be conserved.

A little later Bena, Roiban and Polchinski[ 23] derived the Lax representation
for the $AdS_{5}\times S_{5}$ model. That guarantees the classical
integrability but leaves open the question of possible quantum anomalies.
Recently I found a very simple the Lax representation for the non-critical
cases $AdS_{4}$ (describing 3d gauge theory) and $AdS_{5}\times S_{1}$ ; it is
desirable to study possible anomalies in more details although the above
argument suggests that they must be absent.

This integrability has nothing to do with the supersymmetry of the models. But
is it restricted to the models which are conformal in space-time? I don't
think it is, but more work has to be done. As a first step, I looked at the
integrability conditions for the (classical) sigma models with the target
space warped by the general factor $a^{2}(\varphi).$The method is a direct
generalization of [ 22]. Namely we search for a spin 4 tensor of the form (we
use the light cone coordinates on the world sheet)
\begin{equation}
\Theta_{+}^{(4)}=(\partial_{+}^{2}\varphi)^{2}+f(\varphi)(\partial_{+}%
^{2}x)^{2}+h(\varphi)(\partial_{+}\varphi)^{4}%
\end{equation}
which satisfies the continuity equation%
\begin{equation}
\partial_{-}\Theta_{+}^{(4)}=\partial_{+}S_{+}^{(2)}%
\end{equation}
as a consequence of the equations of motion. It is also convenient to work
with the solutions with zero energy- momentum tensor. One gets somewhat
complicated equations for $a^{2},$all I can say at the moment is that there
are some non-conformal solutions, but much more work has to be done. Still, I
think it is important to pose this well defined mathematical problem.

We see that integrability reveals itself in the three ways. First there are
(perhaps) conserved currents in the loop space (22 ). Then, the spectrum of
the space-time anomalous dimensions is related to the ferromagnetic spin
chains. And finally the string sigma model are integrable and , as typical in
these cases, are (perhaps) related to the antiferromagnetic spin chains. How
these facts are related ? I don't have a full answer to this question.

It seems, first of all, that the integrability of the sigma model implies an
infinite set of relations for the Wilson loop. Indeed, the Wilson loop is
nothing but the wave functional of the sigma model. In any integrable system
we have a set of commuting integrals $I_{k}(x_{s},p_{s}).$The wave function
$\Psi$ (or wave functional) satisfies the simultaneous equations $I_{k}%
\Psi=0.$It is tempting to think that the above conserved currents in the loop
space provide just such relations. This remains to be seen.

Second, the ferromagnetic chain defines the anomalous dimensions in space-
time, while the sigma model and antiferromagnetic chain define anomalous
dimensions on the world sheet. The two are related, as explained above, and so
are integrabilities in both cases. Once again, a much more concrete
explanation is desirable and possible.

Recently I found two non-trivial conformal sigma models [ 24] , describing
gauge theories in which the beta function has an isolated zero. In this case
the curvature is not small anymore and thus the supergravity approximation is
not applicable. The sigma models are based on the coset superspaces
$\frac{SU(2,2|2)}{SO(4,1)\times SU(2)}$which has the bosonic part
$AdS_{5}\times S_{1}$ and $\frac{OSP(2|4)}{SO(3,1)\times SO(2)}$ with the
bosonic part $AdS_{4}$. It is not straightforward to identify these gauge
theories since we can't go to the weak coupling.

Judging by the structure of the RR fluxes one can guess that the first model
contains a theta term. If this guess is correct, we can apply this model to
the solution of the strong $CP$ problem. The phase picture in this model may
be similar to that in quantum Hall effect [25] , namely at $\vartheta=\pi$ we
have a conformal theory, while if we start with $\vartheta<\pi$ in the UV
region, it renormalizes to zero in the IR. At the scale of $\Lambda_{QCD}$ we
have $\vartheta_{IR}\leq10^{-9}$ from the neutron dipole moment constraints.
As we increase energy, the effective $\vartheta$ increases as a power of
energy $\vartheta(q)\sim\vartheta_{IR}(\frac{q^{2}}{\Lambda_{QCD}^{2}%
})^{\alpha}$ where $\alpha$ is related to the presently unknown anomalous
dimension of the relevant operator near the fixed point. At some scale
$\Lambda_{CP}$ the $CP$ violation becomes strong. If we assume that
$\Lambda_{CP}$ $\sim\Lambda_{GUT}$ , we could estimate the neutron dipole
moment, but for that we need the value of $\alpha.$I must add, however , that
there is no evidence that the theory we are discussing is the physical QCD,
since the fields of the fixed point theory are not yet identified. This is a
work for the future.

Another field of knowledge on which the gauge /string correspondence can shed
some light is the meaning of geometry at the Planck scales. We see from the
above that the small curvature limit , which is naturally described in terms
of the Einstein equations corresponds to the very large Yang-Mills coupling
(which is hard to handle directly). Conversely, the limit of large curvatures
corresponds to the small gauge couplings. Moreover "geometry" at infinite
curvature is described by free gauge fields! All possible physical information
about it is encoded in the gauge invariant words, like $Tr($ $F^{k}\nabla
^{l}F^{m}...)$ and their correlation function. The conventional space-time
gradually arises as we decrease the curvature (which is defined through the
gauge coupling).

The situation resembles the thermodynamics / statistics correspondence. In
thermodynamics we introduce temperature and entropy by studying heat transfer.
Moreover, we experience heat with our senses (especially in Princeton). This
is analogous to the description and perception of the continuous space-time.
In statistics we realize that entropy is the logarithm of the number of
configurations of molecules and that the description in terms of temperature
has no meaning whatsoever at the molecular scale . This is similar to our
statement that at infinite curvature we must replace space-time with some
abstract correlation functions of gauge-invariant words.

In my opinion,string theory in general may be too ambitious. We know too
little about string dynamics to attack the fundamental questions of the
"right" vacua, hierarchies, to choose between anthropic and misanthropic
principles etc. The lack of control from the experiment makes going astray
almost inevitable. I hope that gauge/string duality somewhat improves the
situation. There we do have some control, both from experiment and from
numerical simulations. Perhaps it will help to restore the mental health of
string theory.

In '98 I wrote [14] :" There are reasons to believe that the above sigma
models with constant curvature are completely integrable. Thus we may hope to
find the complete solution of the gauge fields -strings problem and perhaps
even to discover experimental manifestations of the fifth (Liouville)
dimension." It seems that we are moving in this direction, although at a much
slower pace than I hoped.

This work was partially supported by the NSF grant 0243680. Any opinions,
findings and conclusions or recommendations expressed in this material are
those of the authors and do not necessarily reflect the views of the National
Science Foundation.

\bigskip REFERENCES

[1] A.A. Migdal A. M. Polyakov Sov. Phys. JETP 24 (1967) 91

[2] A.M. Polyakov Sov. Phys. JETP 28 (1969) 533;

Sov. Phys. JETP 30(1970) 151

[3] A. M. Polyakov Sov. Phys. JETP Lett. 12 (1970) 381

[4] A.M. Polyakov Sov. Phys. JETP 32(1971) 296 ;

Sov. Phys. JETP 33 (1971) 850 ; Sov. Phys. JETP 34 (1972) 1177

[5] V.L. Berezinsky Sov. Phys. JETP 34(1972) 610

[6] A.M. Polyakov Phys. Lett 59B (1975)82

[7] G. t'Hooft, in "The Whys of Subnuclear Physics" Erice 1977

[8] S. Mandelstam Phys. Reports 23 (1976) 245

[9] A.A. Belavin et al. Phys. Lett. 59B (1975) 85

[10] G. t'Hooft Phys. Rev. Lett.37 (1976) 8

[11] A. M. Polyakov Phys. Lett. 72B (1978) 477

[12] L. Susskind Phys. Rev. D20(1979) 2610

[13] A.M. Polyakov Nucl. Phys. B164 (1980)171;

Phys. Lett 82B(1979) 247

[14] A.M. Polyakov Nucl. Phys.B (Proc. Suppl.) 68 (1998)1 ;

Int. Journ. Mod. Phys. 14 (1999)645

[15] S.Gubser, Ig. Klebanov, A.M. Polyakov Phys. Lett.B 428 (1998) 105

[16] E. Witten Adv. Theor. Math. Phys. 2 (1998)697

[17] S.Gubser, Ig. Klebanov, A.M. Polyakov Nucl. Phys.

[18] J. Minahan, K. Zarembo JHEP 0303(2003)013

[19] L. Lipatov JETP\ Lett. 59(1994)596

[20] G. Korchemsky, L. Faddeev Phys. Lett B342 (1995) 311

[21] A. Mikhailov, V. Zakharov Sov. Phys. JETP\ 47 (1978) 1017

[22] A. M. Polyakov Phys. Lett. B72 (1977) 224

[23] I. Bena, J. Polchinski , R. Roiban Phys. Rev. D69 (2004)046002

[24] A. M. Polyakov Mod. Phys. Lett. A 19 (2004) 1661

[25] D. E. Khmelnitsky Sov. Phys. JETP Lett. 38 (1983) 552

\end{document}